\documentclass[runningheads]{llncs}
\usepackage[T1]{fontenc}
\usepackage{graphicx}
\usepackage{filecontents}
\usepackage{latexsym}
\usepackage{threeparttable}
\usepackage{enumitem}
\usepackage{float}
\usepackage{lipsum} 
\usepackage[utf8]{inputenc}
\usepackage{float}
\usepackage{setspace}
\usepackage{amsmath} 
\usepackage{booktabs} 
\usepackage[english]{babel}
\usepackage[utf8]{inputenc}
\usepackage{multirow}
\usepackage{microtype}
\usepackage{inconsolata}
\usepackage{graphicx}
\usepackage{xcolor}
\usepackage{subcaption}

\begin{document}

\title{Leveraging Decoder Architectures for Learned Sparse Retrieval} 



\author{Jingfen Qiao\inst{1}\thanks{Corresponding author} \and
        Thong Nguyen\inst{1} \and
        Evangelos Kanoulas\inst{1} \and
        Andrew Yates\inst{1,2}}

\authorrunning{J. Qiao et al.}
\institute{University of Amsterdam, Netherlands\\
\email{\{j.qiao,t.nguyen2,e.kanoulas\}@uva.nl} \and
           Johns Hopkins University, United States\\
\email{andrew.yates@jhu.edu}}

\maketitle
\begin{abstract}
Learned Sparse Retrieval (LSR) has traditionally focused on small-scale encoder-only transformer architectures. With the advent of large-scale pre-trained language models, their capability to generate sparse representations for retrieval tasks across different transformer-based architectures, including encoder-only, decoder-only, and encoder-decoder models, remains largely unexplored. This study investigates the effectiveness of LSR across these architectures, exploring various sparse representation heads and model scales. Our results highlight the limitations of using large language models to create effective sparse representations in zero-shot settings, identifying challenges such as inappropriate term expansions and reduced performance due to the lack of expansion. We find that the encoder-decoder architecture with multi-tokens decoding approach achieves the best performance among the three backbones. While the decoder-only model performs worse than the encoder-only model, it demonstrates the potential to outperform when scaled to a high number of parameters. 


\end{abstract}

\section{Introduction}
Modern LLMs are knowledgeable about a wide range of topics and have achieved remarkable performance in their application to information retrieval~\cite{ma2023finetuning,pradeep2023rankvicuna,sun2023chatgpt}. However, much of the research has focused on using LLMs to generate dense vectors rather than sparse ones \cite{ma2023finetuning,ma2023zeroshot}. Learned sparse retrieval (LSR) utilizes LLMs to encode queries and documents into lexical sparse vectors, i.e.~vectors where most of the elements are zero. Compared to dense retrieval, LSR's lexical representations are more interpretable as each output dimension is aligned with a term in a vocabulary.  In addition, LSR relies on an inverted index for retrieval, which is smaller in size than a vector index (e.g., HNSW~\cite{malkov2018efficient}) employed by dense retrieval. Recent studies have demonstrated that LSRs could efficiently achieve strong first-stage retrieval performance~\cite{lassance2024spladev3,bruch2024efficient}. Its effectiveness, efficiency and transparency make LSR a compelling alternative to dense retrieval in many information retrieval applications.

\begin{figure}[ht]
    \centering 
    \includegraphics[width=8cm]{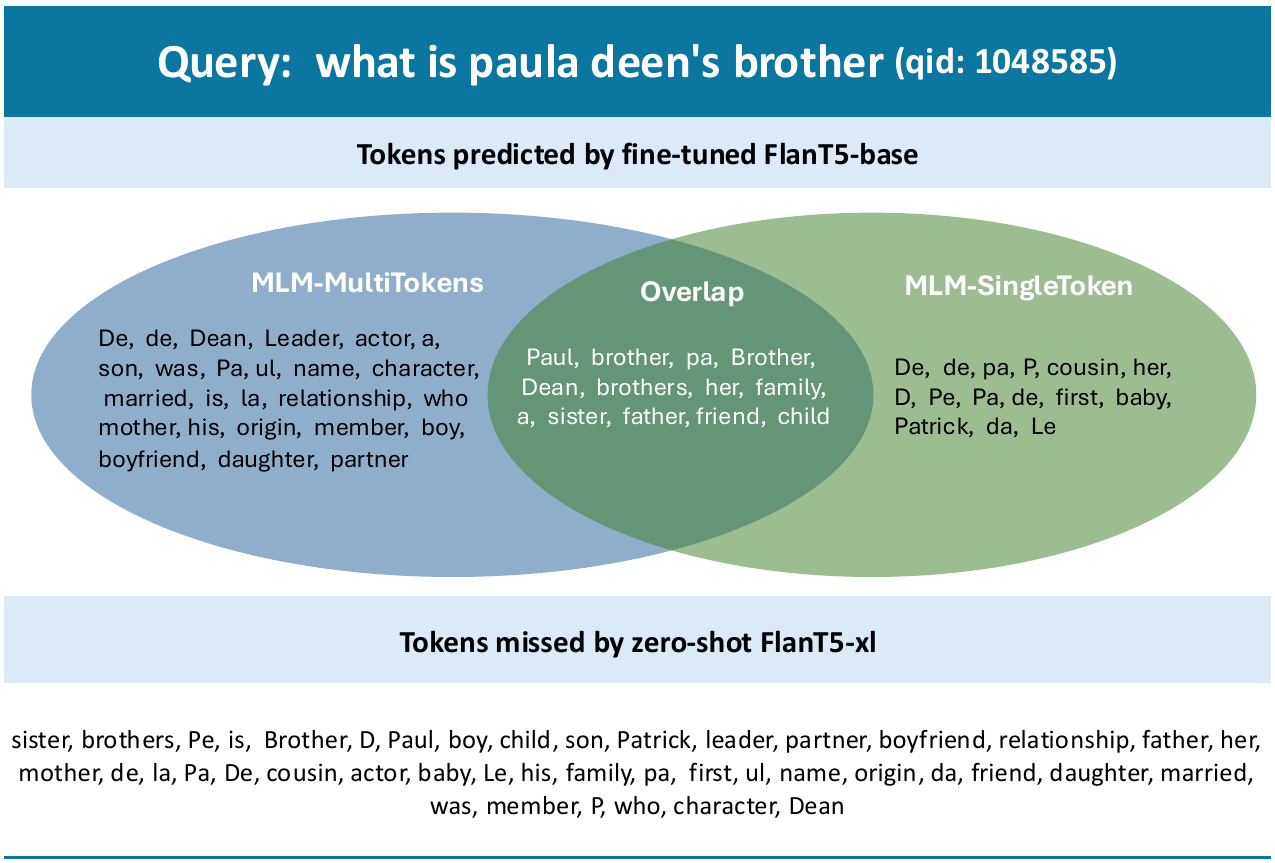}
    \caption{Output Bags of Tokens Produced by Different Sparse Representation Heads; Zero-shot encoder (FlanT5-xl) misses many important expansion terms. The MLM-MultiTokens head captures more relevant tokens than the MLM-SingleToken head by gathering contextual information from all input token rather than a single token. }
    \label{fig: sparse vector examples}
\end{figure}

When evaluating the capabilities of LLMs for sparse retrieval, a key question is whether they can generate effective sparse representations in zero-shot settings. Our analysis shows that while LLMs excel in many generative tasks, they struggle with zero-shot sparse retrieval. As illustrated in Figure \ref{fig: sparse vector examples}, which shows the tokens predicted by different sparse representation heads with a zero shot or fine-tuned model, LLMs not only fail to expand the input text to include semantically relevant terms, but they often introduce noise, leading to semantic drift and poor retrieval performance. This limitation highlights the need for supervised training to enhance LLMs' effectiveness and raises the question of whether fine-tuning could enable LLMs to outperform smaller-scale learned sparse retrieval (LSR) methods.

\begin{figure*}[ht]
    \centering
    \includegraphics[width=13cm]{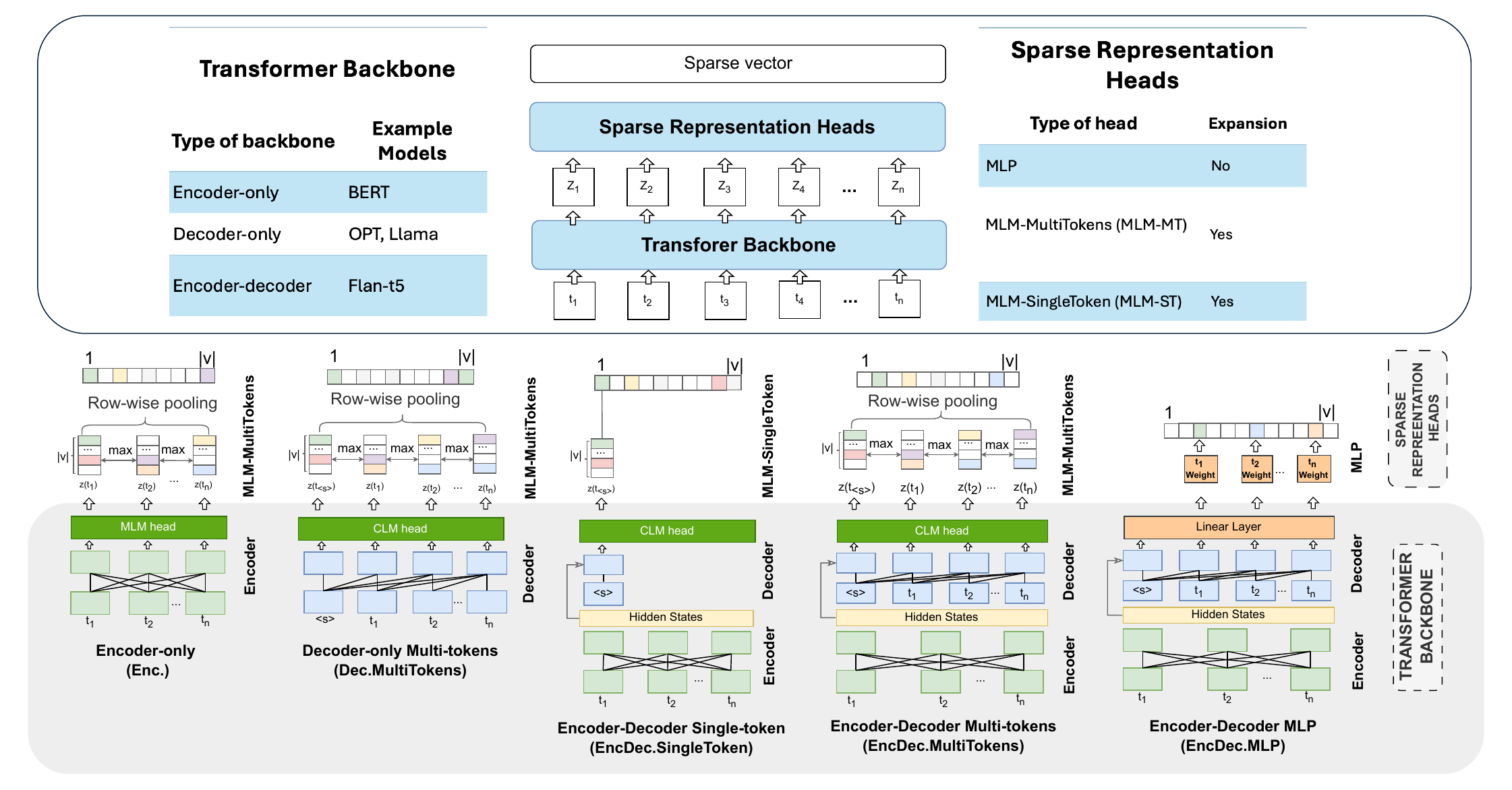}
    \caption{Learned sparse retrieval architectures consist of (1) a transformer backbone that takes query or document text as input and outputs hidden state(s) and (2) a sparse representation head that takes the hidden state(s) as input and outputs sparse lexical representations. 
    }
\label{fig:architectures}
\end{figure*}

Previous research on LSR has predominantly focused on small-scale encoder-only transformer architectures~\cite{bai2020sparterm,formal2021splade,MacAvaney_2020,paria2020minimizing,10.1145/3269206.3271800}. For instance, Splade~\cite{formal2021splade}, a state-of-the-art LSR model, is built upon Cocondenser~\cite{gao2021unsupervised}, which utilizes a BERT-base architecture with 110M parameters, specifically pre-trained for Information Retrieval (IR) tasks. The generalizability of LSR to other transformer architectures, such as decoder-only and encoder-decoder, remains unclear, as does its scalability concerning model size. 
Decoders are trained to sequentially generate a subsequent token based on the given input text and previously generated tokens. This is different from sparse representation learning which requires weighting existing tokens within the input and expanding them with contextually relevant terms. To learn a sparse representation from the decoder output, we investigate using a multi-tokens decoding approach. 
This adaptation allows the decoder to work more like an encoder, making better use of the parameters within the decoder module. Our results indicate that the multi-tokens decoding approach for encoder-decoder models helps to effectively utilize the parameters in both encoder and decoder modules.

This paper aims to address these gaps by investigating LSR across different transformer architectures (encoder-only, decoder-only, and encoder-decoder) at several scales and exploring the capabilities of LLMs under zero-shot and fined-tuning settings. Our findings are as follows: \textbf{(1)} Without fine-tuning, LLMs struggle to generate meaningful sparse representations due to inappropriate term expansions; \textbf{(2)} the ability of decoder-only architectures to produce sparse representations is limited and comparable results to encoder-only models are only achieved when the model is scaled to a large size, such as 3 billion parameters; \textbf{(3)} by copying the encoder's input to the decoder we enable the encoder-decoder to perform multi-tokens decoding effectively aggregating lexical information across the input sequence and creating a significantly more effective sparse representation compared to the single-token approach; and \textbf{(4)} encoder-decoder architectures with multi-tokens decoding can outperform encoder-only and decoder-only.\footnote{Code: https://github.com/JingfenQiao/Decoder4LSR}

\section{LSR Architectures}
In LSR, queries and passages are encoded into sparse representations over terms (Figure~\ref{fig:architectures}). In this section, we describe how to generate sparse representations from various transformer-based architectures, including encoder-only, decoder-only, and encoder-decoder. 

\subsection{Transformer Backbones}
Given a sequence of \(n\) tokens \((t_1, t_2, \ldots, t_n)\) in the query or passage, the task of a LSR model is to predict term weights based on the hidden states output by the transformer-based backbone. Therefore, the transformer backbone aims to transform each input token \(t_i\) into a hidden state (latent vector) $h_i$. A sparse representation of the input is then produced by passing these hidden states to a sparse representation head. In this section, we first describe how the hidden states $h_i$ are produced by encoder-only backbone in the previous research, and then illustrate how to leverage decoders to construct sparse representations.

We will use the following notation to describe each transformer-based backbone in this paper: \(\oplus\) denotes the concatenation operator, \(h_{1:n}\) denotes the hidden states corresponding to input tokens \(t_{1:n}\) that are produced by a transformer-backbone, \(v_i\) and \(e_i\) represents the \(i_{th}\) token and transformer’s initial input embedding in the model vocabulary \(|V|\), where \(i \in \{0, \dots, |V|-1\}\).\\

\noindent \textbf{Encoder-only (Enc.)}: When using a transformer encoder stack as the backbone, we feed text directly into the encoder, which produces a contextualized embedding of each input term. We refer to these as hidden states $h_i$ for consistency with the decoder-based backbone that we introduce in this paper.
    \begin{equation}
        \small
        h_{1:n} = \textrm{Encoder}(t_1, t_2, \ldots, t_n)
        \label{eq: 1}
    \end{equation}
When using an encoder backbone, the hidden states are produced using bi-directional attention that considers terms both before and after a given term $t_i$ when producing hidden states to represent the input sequence. This is an intuitive property for the transformer backbone to have. With the exception of the zero-shot PromptRep model~\cite{zhuang2024promptreps} and orthogonal document expansion approaches like doc2query \cite{nogueira2019document}, all existing learned sparse retrieval models use backbones based on a transformer encoder~\cite{nguyen2023unified}.\\

\noindent \textbf{Decoder-only with multi-tokens (Dec. MultiTokens)}: In contrast to encoder models, decoders are designed with causal attention, where the generation of a hidden state $h_i$ depends only on the previously generated tokens, and the decoder starts with the start symbol of a model \texttt{<s>}. We concatenate the start symbol of the model \texttt{<s>} in front of the token \(t_1\).

\begin{equation}
    \small
        h_{1:n} = \textrm{Decoder}(\textrm{<s>}, t_1, t_2, \ldots, t_n)
    \label{eq: 2}
\end{equation}

When using decoder-based backbones, rather than invoking the decoder repeatedly for each token, we feed the entire input text to the decoder in a single pass to generate sparse representations. This contrasts with the typical token-by-token decoding approach, where a new hidden state is produced at each decoding step. As illustrated in Figure~\ref{fig: architectures}, the row-wise pooling of MLM outputs demonstrates how the hidden state of every token contributes to the final multi-token representation.

\noindent \textbf{Encoder-Decoder with single-token (EncDec. SingleToken)}: In encoder-decoder with single-token, similar to the Sentence-T5 embedding model~\cite{ni2021sentencet5}, input text is fed into the encoder and only the start token \texttt{<s>} is fed into the decoder. Therefore, it only yields a single latent vector $h_1$ that is derived from the hidden state of the input token \texttt{<s>}, which is in turn derived from the hidden states of the entire input text produced by the encoder stack.
\begin{equation}
        \small
        h^{enc}_{1:n} = \textrm{Encoder}(t_1, t_2, \ldots, t_n) 
        \label{eq: 3}
\end{equation}
\begin{equation}
        \small
        h_1 = \textrm{Decoder}(h^{enc}_{1:n} \oplus e_{<s>})_{n+1}
        \label{eq: 4}
\end{equation}
    
\noindent \textbf{Encoder-Decoder with multi-tokens (EncDec. MultiTokens)}: We hypothesize that the output representation of the encoder-decoder architecture is bottlenecked by the single decoder input, the start token \texttt{<s>}, which limits the model's ability to aggregate lexical information from the entire input sequence. To overcome this bottleneck, we propose a multi-tokens decoding approach that allows the model to gather information from any input token.  As shown in Figure~\ref{fig: architectures}, the Encoder-Decoder with multi-tokens model receives full input text both in the encoder and decoder, and the decoder’s input is prepended with the start token.

\begin{equation}
    \small
    h^{enc}_{1:n} = \textrm{Encoder}(t_1, t_2, \ldots, t_n)
    \label{eq: 5}
\end{equation}
\begin{equation}
    \small
    h_{1:n} = \textrm{Decoder}(h^{enc}_{1:n}\oplus e_{<s>} \oplus t_1, t_2, \ldots, t_n)
\label{eq: 6}
\end{equation}

\subsection{Sparse Representation Heads}
A sparse representation head takes a hidden state produced by the transformer backbone as input and produces one or more term weights.\footnote{Nguyen et al.~\cite{nguyen2023unified} refer to these as \textit{sparse encoders}. We use the term \textit{head} instead to avoid overloading the term \textit{encoder}, which can also refer to the type of transformer backbone used.} These term weights are then aggregated to produce a sparse representation of the entire input text, which can then be used for retrieval by taking the dot product of a query representation and a document representation. More formally, a sparse representation head takes a sequence of of $n$ hidden states output by a transformer backbone \(h_{1:n} = (h_1, h_2, \ldots, h_n)\) as input. The head outputs a sparse vector with $|V|$ dimensions where $v_i$ corresponds to a term $t_i$ in the vocabulary.  

\textbf{MLP head: } The MLP head, used in models like DeepCT~\cite{hofstätter2021improving}, uniCOIL~\cite{lin2021brief} and DeepImpact~\cite{mallia2021learning}, uses a multi-layer perceptron to process contextualized embeddings for each input term in the query or passage. It assigns non-zero weights to terms that occur only in the input text. The importance \(MLP(h)_i\) of each term \(v_i\) within a model vocabulary \(V\) for a given input \(t_n\) is defined by Equation \ref{eq: 7}, where \(W\) and \(b\) are the weight and bias of the linear head, respectively. \(1(v_i = t_n)\) ensures to that only tokens in the input receive weights, the other terms are zero. 
\begin{equation}
    \small
    MLP(h_{1:n})_i = \sum_{j=1}^{n} \log \left(1(v_i = t_j) \cdot \text{ReLU}(h_j W + b)  + 1 \right)
    \label{eq: 7}
\end{equation}

\textbf{MLM head: } In contrast to the MLP head, the MLM head expands the input (using the model's vocabulary) in addition to weighting tokens. MLM heads have been used in models such as SPLADE~\cite{formal2021splade}, Sparta~\cite{zhao2020sparta}, and TILDE~\cite{TILDE}. The MLM head uses the logits value produced by the language model head for each input term \(v_i\) to produce a sparse representation. It can assign a weight to to any terms in the model vocabulary \(V\) regardless of whether it appears in the input text, thereby exploiting the language model's capability to expand and weight terms. The importance \(MLM(h_{1:n})_i\) of each term \(v_i\) within a model's vocabulary \(V\) is aggregated from the logits matrix along the input sequence, as defined in Equation \ref{eq: 8}. The MLM-SingleToken (MLM-ST) process a single hidden state \(h_1\) from the transformer backbone to determine the importance of each term in the vocabulary, while MLM-MultiTokens (MLM-MT) processes multiple hidden states \(h_{1:n}\) and takes the max weight over the different hidden states.

\begin{equation}
    \small
    MLM(h_{1:n})_i = \max_{j=1}^{n}\log \left( 1 + \text{ReLu} \left( h_j^T e_i + b_i \right) \right)
    \label{eq: 8}
\end{equation}

The \(ReLu\) activation function and logarithmic normalization are applied to ensure the positive of term weights~\cite{formal2021splade} and prevent some terms from dominating in the sparse representation~\cite{fang2004formal}. The max-pooling operator in Equation \ref{eq: 8} is to aggregate term importance weights into a document-level sparse representation. Formal et al.,~\cite{formal2021splade} found that max pooling is more effective than sum pooling. Therefore, we employed max-pooling for all MLM models.

\section{Experimental Setup}
In this section, we outline the experimental design, including the models, training configurations, and evaluation methods, to investigate learned sparse retrieval (LSR) performance across various transformer architectures.

\textbf{Retrieval and Indexing:} We apply the same architecture type with shared weights for both the query and document encoders. For indexing and retrieval, we use the Anserini toolkit~\cite{10.1145/3077136.3080721,10.1145/3239571} to index the encoded documents and retrieve with the encoded queries.

\textbf{Dataset and Evaluation:} We trained our models on the MS MARCO passage dataset~\cite{bajaj2018ms}, which includes 8.8 million passages and 40 million training triplets. All models are trained on those triplets and hard negatives generated by the cross-encoder teacher model in~\cite{hofstätter2021improving}. We report the MRR@10, NDCG@10 and Recall@1000 of all models on the MS MARCO Dev and TREC Deep Learning (TREC-DL) 2019/2020~\cite{craswell2021overview,craswell2020overview}. In addition of the effectiveness metrics, we report the FLOPs metric (the average number of term overlap between a pair of query and document) as a proxy of efficiency.

\textbf{Backbones and baselines} We selected three different pretrained transformer language model as the backbone of our study, including encoder-only (Distilbert-base-uncased~\cite{Sanh2019DistilBERTAD}), decoder-only (OPT~\cite{zhang2022opt}) and encoder-decoder (T5~\cite{chung2022scaling}). Distilbert-base-uncased is widely recognized and extensively employed within the LSR literature~\cite{doshi2024mistralspladellmsbetterlearned,formal2021splade,lassance2024spladev3}. The OPT~\cite{zhang2022opt} and Flan-T5~\cite{chung2022scaling} offers a wide range of parameter sizes. This variability enables us to explore how decoder-only architectures scale and perform in generating sparse representations, providing insights into their effectiveness across different model sizes.  In terms of baselines, we select based on query expansion and term weighting strategies and compare against the unsupervised sparse model BM25 and the supervised sparse model, including EPIC~\cite{macavaney2020expansion}, DeepCT~\cite{dai2019contextaware}, DocT5Query~\cite{nogueira2019document}, DeepImpact~\cite{mallia2021learning}, Splade-max~\cite{formal2021splade} and DistilSplade-max~\cite{formal2021splade}.  The unsupervised sparse baseline BM25 employs neither expansion nor supervised weighting. Supervised LSR baselines include EPIC~\cite{macavaney2020expansion} and DeepCT~\cite{dai2019contextaware}, which use supervised weighting without query expansion; DocT5Query~\cite{nogueira2019document}, which focuses solely on document term expansion without term weighting; DeepImpact~\cite{mallia2021learning}, which applies supervised term weighting without query expansion; and Splade-max~\cite{formal2021splade} and DistilSplade-max~\cite{formal2021splade}, which integrate both supervised query expansion and term weighting. 

\textbf{Training Configuration} All models were fine-tuned using the ADAM optimizer with $6000$ warmup steps. The batch size for all models was consistently maintained at 16. Due to model nuances, the OPT model uses a lower learning rate of \(1^{e-6}\), while Flan-T5 uses \(5^{e-4}\). We used one NVIDIA A6000 GPU with 48GB memory to train medium-sized models, such as Flan-T5-base, and OPT-3.5. Larger models (OPT-1.3B, OPT-2.7B, Flan-T5-large and Flan-T5-xl) were trained on four A6000 GPUs. Following the setting of Splade~\cite{formal2021splade}, we schedule the regularization weight \(\lambda\) to increase quadratically over the first 100k steps and we tried different values of $\lambda$  ranging from \(1^{-1}\) to \(1^{-4}\).

\section{Optimization}

Knowledge distillation has emerged as a highly effective strategy for transferring knowledge from a high-performing cross-encoder to a more efficient bi-encoder in information retrieval tasks~\cite{formal2021splade,lassance2024spladev3,hofstätter2021improving}. By training the student model to replicate the teacher’s score distribution, the student preserves a significant portion of the teacher’s ranking ability. Compared to alternative approaches for retrieval, such as contrastive learning or reinforcement learning, the knowledge distillation inherits a well-calibrated signal from a strong teacher model, thus reducing the need for labor intensive training sample construction or careful reward engineering. Concretely, given a batch \(B\) of training triplets \(q_i, P^+_i, P^-_i\) consisting of a query \(q_i\), a positive document \(P^+_i\) and a negative document \(P^-_i\), we compute the margin between the positive and negative documents for both teacher and student. We then minimize the mean squared error (MSE) of the difference between these margins, as shown in Equation \ref{eq: 9}. 

\begin{equation}
    \mathcal{L}_{\text{MarginMSE}} = \frac{1}{|B|} \sum_{i = \{0,\dots,|B|-1\}} \left( \Delta M_s^i - \Delta M_t^i \right)^2
    \label{eq: 9}
\end{equation}

where \(\Delta M_s^i = S(q_i, P^+_i) - S(q_i, P^-_i)\) and \(\Delta M_t^i = T(q_i, P^+_i) - T(q_i, P^-_i)\) are the respective margin scores of the student \(S\) and teacher model \(T\) on the i-th triplet, and 
\(|B|\) is the batch size. Following~\cite{hofstätter2021improving,formal2021splade}, we use MarginMSE as the distillation loss to ensure fair comparisons and reduce confounding factors. Although a more complex training setup (e.g. using an ensemble of multiple teachers, dual MarginMSE and KL-Div loss, continuous training as in Splade-V3~\cite{lassance2024spladev3}) could be applied to increase the overall effectiveness, we believe that our relative findings between different architectures and approaches do not change.

\textbf{Distillation Training Data} To investigate how the scaling of teacher models affects performance, we experiment with two teachers: MiniLM-L-6-v2~\cite{wang2020minilmdeepselfattentiondistillation} and Rankllama-13B~\cite{ma2023finetuning}. Rankllama-13B outperforms MiniLM-L-6-v2 on the MSMARCO training set, as shown in Table~\ref{tab: table 5} in Appendix. For both teacher models, we collect hard negatives from multiple retrieval systems (including dense retrieval models and BM25) to provide challenging negative samples. Each teacher then generates score pairs for the query-positive and query-hard-negative pairs using the hard negatives from Sentence-Transformers\footnote{https://huggingface.co/datasets/sentence-transformers/msmarco-hard-negatives}. We use the smaller teacher’s scores from~\cite{formal2022distillation} directly, whereas we generate the Rankllama-13B scores ourselves\footnote{https://huggingface.co/datasets/lsr42/rankllama-ms-marco-scores}. We observe that Rankllama-13B produces a relatively narrow score distribution (Appendix Table~\ref{fig: 3}), which may make it more difficult for students to learn effectively. This aligns with findings by \cite{lassance2024spladev3}, who show that the MarginMSE-based distillation process is sensitive to score distributions. To address this, we apply a affine transformation to flatten Rankllama-13B’s scores, ensuring that its mean and standard deviation more closely match those of MiniLM-L-6-v2.

\textbf{Sparsity Regularization} Finally, to encourage sparsity in the learned representations, we adopt FLOPs regularization~\cite{paria2020minimizing} and activation functions ~\cite{Arpit2015WhyRA} to facilitate the learning of distributed sparse embeddings. FLOPs are a differentiable relaxed approximation of the number of floating-point operations(FLOPs) to compute the score between query and document. Formally, our overall objective is defined in Equation \ref{eq: 10}.

\begin{equation}
    \small
    \begin{split}
    \mathcal{L}_{\text{ranking}} = \mathcal{L}_{\text{MarginMSE}} + \lambda_{q}\mathcal{L}_{reg}^{q}  + \lambda_{d}\mathcal{L}_{reg}^{d}
    \end{split}
    \label{eq: 10}
\end{equation}

where \(\mathcal{L}_{reg}^{q}\) and \(\mathcal{L}_{reg}^{d}\) denote FLOPs-based regularization terms for query and document representations, respectively, and \( \lambda_{q}\) and \( \lambda_{d}\) are their weighting coefficients.

\begin{table*}[ht]
  \caption{Comparison of sparse retrieval without fine-tuning on MS MARCO dev. \(\times\) results are from \cite{mallia2021learning}; A \(\dagger\) indicates paired t-test \(p < 0.05\). (\(\dagger\) indicates test between term expansion and non-expansion.}
    \label{tab: table 2}
      \centering
    \begin{tabular}{lllcl}
         \toprule
         \toprule
        \textbf{Model}  & \textbf{Weighting} & \textbf{Expansion} & \textbf{RR@10} & \textbf{R@1k} \\
          \cmidrule(r){1-5}
       \multicolumn{5}{c}{\textbf{Sparse retrieval}}\\
        (1) BM25$^{\times}$ & Yes&No& 18.8 & 85.8\\
      (2) BM25+Doc2Query$^{\times}$ & Yes&Yes& 27.8 &  94.7\\
      \cmidrule(r){1-5}
      \multicolumn{5}{c}{\textbf{Learned sparse retrieval (zero-shot)}}\\
    (3) EncDec.MultiTokens$_{\text{FlanT5-xl}}$& Yes& Yes & \phantom{0}1.30 & 16.5\\
    (4) EncDec.MultiTokens$_{\text{FlanT5-xl}}$ & Yes & No & \phantom{0}3.67$^{\dagger}$ & 34.0$^{\dagger}$\\
    (5) Dec.MultiTokens$_{\text{OPT-2.7B}}$  & Yes&No& 13.7& 71.4 \\
    (6) Dec.MultiTokens$_{\text{OPT-6.7B}}$  & Yes&No& 12.5&70.5 \\
     \bottomrule
     \bottomrule
    \end{tabular}
\end{table*}

\section{Results and Discussion}
\textbf{RQ1: Can LLMs effectively generate sparse representations in a zero-shot setting when prompted?} \\
Term weighting and expansion are key factors in the effectiveness of sparse retrieval systems~\cite{lin2021brief}. We evaluate LLMs for sparse vector creation with and without term expansion. Recently,  Zhuang et al.~\cite{zhuang2024promptreps} proposed generating sparse and dense representations by prompting LLMs, using the logits of the last token in the input sequence. Their method only assigns weights to tokens present in the input, without term expansion. To investigate LLM's term expansion ability, we compare sparse vectors with and without expansion. In the expansion case, we keep the top 1,000 tokens, while the non-expansion approach uses only tokens from the input text. Instead of relying on the last token’s logits, our method aggregates logits from all input tokens, following Equation~\ref{eq: 8}.


Table \ref{tab: table 2} shows the zero-shot performance of LLMs versus traditional methods like BM25 and BM25+Doc2Query. All zero-shot models use the MLM-MultiTokens head for aggregating term weights. We find that enabling term expansion in Flan-T5-xl leads to a significant drop in Recall@1k (34.0 to 16.5), as noisy terms like 'ray,' 's,' and 'mil' receive high weights, harming performance. OPT-2.7B achieves the best results among LLMs (MRR 13.7, Recall@1k 71.4) but still falls short of BM25 and state-of-the-art LSR models, highlighting the need for fine-tuning to improve term importance assessment in LSR. Additionally, we assessed their effectiveness across various model sizes to provide a more comprehensive comparison, as shown in Table \ref{tab: table 6} in the Appendix.\\

\textbf{RQ2: Can encoder-decoder or decoder-only backbones outperform encoder-only backbones when using multi-tokens decoding approach?} \\

\begin{table*}[ht]
  \caption{Evaluation of different transformer backbones on MS MARCO passage (dev set) and TREC DL 2019/2020. \textit{All models are trained with 600k steps and use the MLM-MultiTokens head for both query and document encoding. \(\ast\) indicates reproduced results.}} 
  \label{tab: table 1}
  \renewcommand{\arraystretch}{1.3} 
   \resizebox{\linewidth}{!}{%
   \begin{tabular}{llcccccc}
   \toprule
    \toprule
     & &&  \textbf{DL2020} & \textbf{DL2019} &\multicolumn{3}{c}{\textbf{MS MARCO dev}} \\
    \cmidrule(lr){4-8} 
    \textbf{Models}  & \textbf{Backbones} & \textbf{FLOPs} &  \textbf{nDCG@10} &  \textbf{nDCG@10} & \textbf{nDCG@10} & \textbf{R@1k} & \textbf{RR@10} \\
    \midrule
    (1) Splade-max$^{\ast}$ \small \cite{formal2021splade} & Encoder-only & \phantom{0}1.3& 67.0 & 68.0 &40.2 & 96.5 & 34.0 \\
    (2) DistilSplade-max$^{\ast}$ \small \cite{formal2021splade} & Encoder-only & 
    \phantom{0}4.0& 67.9 & 71.0 &43.3 & 97.9 & 36.8 \\  
    (3) Enc.$_{\text{FlanT5-base}}$ &  Encoder-only & 16.0 & 61.0&  65.0 &36.0 & 96.9 & 29.8 \\
    (4) Dec.MultiTokens$_{\text{FlanT5-base}}$ & Decoder-only & \phantom{0}2.3 & 67.4 & 68.7 &39.9 & 97.1 & 33.6 \\
    (5) Dec.MultiTokens$_{\text{OPT-350m}}$ & Decoder-only & \phantom{0}4.9 & 68.6 & 66.4  & 39.8 & 97.2 &33.7\\
    (6) EncDec.MultiTokens$_{\text{FlanT5-base}}$ & Encoder-Decoder  & \phantom{0}2.8 & 70.5 & 71.4 &43.6 & 98.3 & 36.8 \\
     \bottomrule
     \bottomrule
  \end{tabular}
  }
\end{table*}

\textbf{Encoder-only VS Encoder-decoder backbone }
The encoder-decoder backbone generally surpasses the encoder-only backbone, as shown in Table \ref{tab: table 1}. Specifically, EncDec.MultiTokens (row 6) outperforms the encoder-only model DistilSplade-max (row 2) in NDCG@10 on the MS MARCO development set, with lower FLOPs (4 vs. 2.8) using the same training configuration. Upon inspection, we find that the encoder-decoder model is more effective at expanding the input queries and documents to relevant terms.
The encoder-decoder architecture benefits from a hybrid attention mechanism, which includes bidirectional attention in the encoder and causal attention in the decoder. This allows it to identify different patterns of semantic dependency in the input. Interestingly, the encoder-only model using Flan-T5 (Enc.$_{\text{FlanT5-base}}$) performs significantly worse than the encoder-only model DistilSplade-max, especially on the MRR@10 and nDCG@10 metrics. This underperformance may be attributed to the fact that Flan-T5 was pre-trained with an encoder-decoder architecture, where the MLM head is only attached to the decoder's hidden states during training. When we attach the same MLM head to the encoder, there may be an incompatibility that makes LSR training more challenging.

\begin{table*}
  \caption{Performance of different Sparse Representation Heads on MS MARCO passage (dev) and TREC DL 2019/2020.   
  \textit{All models are trained with 600k steps. \(\ast\) indicates reproduced results; \(\times\) results are from the work in \cite{mallia2021learning}; \(\times \times\) results are from the work in \cite{nguyen2023unified}; A \(\dagger\) indicates paired t-test \(p < 0.05\). (\(\dagger\) indicates test between MLM-MT and MLM-ST);  A \(\ddagger \) indicates paired t-test \(p < 0.01\)( \(\ddagger\) indicates test between models where both query and passage use MLM-MT and models where either query or passage use MLM-MT)}} 
  \label{tab: table 3}
  
  \renewcommand{\arraystretch}{1.4} 
     \resizebox{\linewidth}{!}{%
    \Large
    \begin{tabular}{llllccccc}
    \toprule
    \toprule
     &  &  &  & \textbf{DL2020} &  \textbf{DL2019} &\multicolumn{3}{c}{\textbf{MS MARCO dev}} \\
     \cmidrule{5-9}
    \textbf{Models}  &  \textbf{Query} & \textbf{Passage} &\textbf{FLOPs} &  \textbf{nDCG@10} &  \textbf{nDCG@10} & \textbf{nDCG@10} & \textbf{R@1k} & \textbf{RR@10} \\
    \midrule
    (1) BM25$^{\times}$ & & & - & 48.7 &  49.7 & 23.5 & 85.8 & 18.8 \\
    (2) EPIC$_{top400}^{\times\times}${\small ~\cite{macavaney2020expansion}}  & MLP & MLM-MT & - & 71.8 & 70.9 & - & 97.2 & 37.2\\
    (3) DeepCT$^{\times}${\small\cite{dai2019contextaware}} &  MLP&MLP &-& 55.0 &  57.8 & 29.8 & 91.0 & 24.4 \\
    (4) DocT5Query$^{\times}${\small\cite{nogueira2019document}} &  - & - &-& 61.9 &  64.8 &33.8 & 94.7 & 27.8 \\
    (5) DeepImpact$^{\times}${\small\cite{mallia2021learning}} &  BINARY&expMLP&-& 65.1 &  69.5 &38.5 & 94.8 & 32.6 \\
    (6) Splade-max$^{\ast}${\small\cite{formal2021splade}} &  MLM-MT & MLM-MT & 1.3 & 67.0 & 68.0 &40.2 & 96.5 & 34.0 \\
    (7) DistilSplade-max$^{\ast}${\small\cite{formal2021splade}} &  MLM-MT&MLM-MT& 4.0  & 67.9 & 71.0 &43.3 & 97.9 & 36.8 \\
    \midrule 
    (8) EncDec.MultiTokens$_{\text{FlanT5-base}}$ & MLP&MLM-MT & 1.3 &  53.6$^{\ddagger}$ &  56.6$^{\ddagger}$ & 41.7$^{\ddagger}$ & 97.6$^{\ddagger}$ & 35.3$^{\ddagger}$\\
    (9) EncDec.MultiTokens$_{\text{FlanT5-base}}$ &  MLM-MT&MLP& 3.3 &50.6$^{\ddagger}$  & 48.1$^{\ddagger}$ & 31.3$^{\ddagger}$ & 95.9$^{\ddagger}$ & 25.8$^{\ddagger}$\\
    (10) EncDec.SingleToken$_{\text{FlanT5-base}}$ &  MLM-ST&MLM-ST&2.5& 54.5 &  59.7 & 36.1 & 95.0 & 30.4 \\
    (11) EncDec.MultiTokens$_{\text{FlanT5-base}}$
    &  MLM-MT&MLM-MT&\textbf{2.8}& \textbf{70.5}$^{\dagger}$ & \textbf{71.4}$^{\dagger}$ &\textbf{43.6}$^{\dagger}$ & \textbf{98.3}$^{\dagger}$ & 36.8$^{\dagger}$ \\
     \bottomrule
     \bottomrule
      \end{tabular}
      }

\end{table*}

\textbf{Decoder-only backbone effectiveness} We trained different LSR variants using the Flan-T5-base (248 million parameters) and OPT-350M (350 million parameters) decoder-only checkpoints. Among all the backbone variations (rows 3, 4, and 6) of Flan-T5-base, the decoder-only variant (row 4) was the second most effective, trailing only behind the encoder-decoder architecture (row 6). Furthermore, compared to DistilSplade-max (row 2), which is built on the BERT encoder, the decoder-only model using the OPT-350M checkpoint (row 5) shows a slight improvement in terms of NDCG@10 and MRR@10 metrics on the DL2020 dataset, but not on MS MARCO and DL2019. Additional insight could also be seen on Table \ref{tab: table 4}, where we observe that the effectiveness of the Decoder-only model (row 8) can exceed Enc. (row 1) on DL2020, DL2019 and MS MARCO when the parameters are scaled to 1.3 billion. \\

\textbf{RQ3: Which sparse representation head is better for creating a sparse representation?} \\

\begin{table}[ht]
  \caption{Impact of scaling student and teacher model. \textit{All models are trained with 600k steps and use the MLM-MultiTokens head for both querying and document encoding. A \(\dagger\)  \(\ddagger\) indicates a paired t-test \(p < 0.05\). (\(\dagger\) indicates a test between models tuned with the teacher MiniLM-L-6-v2 and Rankllama-13b. \(\ddagger\) indicates between smaller student and larger student LSR model with same backbones)}}
  \label{tab: table 4}
    \centering
    \begin{tabular}{lcccc}
    \toprule
    \toprule
      & \textbf{DL2020} & \textbf{DL2019}  & \multicolumn{2}{c}{\textbf{MS MARCO dev}}\\
     \cmidrule{2-5}
    \textbf{Models}  & \textbf{nDCG@10} & \textbf{nDCG@10} & \textbf{RR@10} & \textbf{FLOPs}\\
    \midrule
    (1) Enc.$_{\text{Distilbert-base-uncased}}$ & 67.9 & 71.0  & 36.8  & 4.0 \\ 
    (2) EncDec.MultiTokens$_{\fontsize{20}{3}\selectfont \text{FlanT5-base}}$ & 70.5 & 71.4  & 36.8  & 2.8 \\ 
    (3) Dec.MultiTokens$_{\text{OPT-350M}}$ & 68.6 & 66.4  & 33.7 & 4.9 \\ 
    \midrule
    \multicolumn{4}{c}{\textbf{Scaling Teacher}} & \\
    (4) Enc.$_{\text{Distilbert-base-uncased}}$ & 73.1$^\dagger$ & 75.1$^\dagger$  & 37.3  & 5.8 \\ 
    (5) EncDec.MultiTokens$_{\text{FlanT5-base}}$ & 70.1 & 73.4  & 37.8 & 5.2 \\ 
    (6) Dec.MultiTokens$_{\text{OPT-350M}}$ & 61 & 66.8  & 32.8 & 5.1 \\
    \midrule
    \multicolumn{4}{c}{\textbf{Scaling Student}} & \\
    (7)  EncDec.MultiTokens$_{\text{FlanT5-large}}$ & 70.7 & 73.3  & 37.8  & 3.8 \\ 
    (8)  Dec.MultiTokens$_{\text{OPT-1.3B}}$ & 69.5  &73.1$^\ddagger$  & 36.9$^\ddagger$ & 4.8 \\
    \bottomrule
    \bottomrule
  \end{tabular}
\end{table}

\textbf{MLM-MultiTokens (MLM-MT) vs. MLM-SingleToken (MLM-ST)} We evaluate the MLM-MultiTokens and MLM-SingleToken heads on the best-performing encoder-decoder model using the Flan-T5 base checkpoint. As shown in Table \ref{tab: table 3}, the MLM-MultiTokens head outperforms the MLM-SingleToken head across all evaluated metrics and datasets (MS MARCO, DL2019, DL2020). This suggests that the MLM-MultiTokens head, which aggregates latent vectors from multiple input tokens, facilitates more effective sparse representations compared to the single token approach from the decoder output.

\textbf{MLM-MultiTokens vs. MLP} We also compare the performance of the MLP and MLM-MultiTokens heads derived from the decoder output using the Flan-T5-base checkpoint. The results in Table \ref{tab: table 3} show that encoding both the query and document with the MLM-MultiTokens head achieves the highest effectiveness, significantly outperforming the MLP head, particularly in terms of nDCG@10 across different datasets. This contrasts with the results observed on an encoder-only backbone, where switching the query encoder from MLM-MultiTokens to MLP—reproduced using hard negative distillation labels in the study by~\cite{nguyen2023unified}—did not yield a significant improvement. These findings suggest that MLM-MultiTokens head query expansion is more beneficial for encoder-decoder backbones compared to encoder-only backbones.\\

\textbf{RQ4: How is the performance of the LSR affected by scaling the teacher and student models on the different backbones?}

Table~\ref{tab: table 4} illustrates the impact of scaling teacher and student models across different backbones, including Encoder-only, Encoder-decoder, and Decoder-only.

\textbf{Impact of scaling teacher model} To assess the impact of scaling the teacher model on different backbones, we employed a more effective cross-encoder RankLLaMA-13B~\cite{ma2023finetuning} as the teacher to supervise LSR model (student) tuning.  Row 4-6 shows the effect of using a larger teacher model across different backbones. Consistent with findings by \cite{lassance2024spladev3}, the encoder-only model shows significant improvements when scaling the teacher model, increasing from 67.9 to 73.1 on DL2020 and from 71.0 to 75.1 on DL2019. The encoder-decoder backbone also exhibits improvements on DL2019 and MS MARCO, though to a lesser degree. In contrast, the decoder-only model experiences a decline in performance on MS MARCO and DL2020. This decline may be attributed to the sharp distribution of RankLlama teacher model scores, as shown in Figure \ref{fig: 3}. The sharp distribution makes it challenging for the decoder-only architecture to generate a sparse vector by relying solely on FLOPs regulation in the loss function. Further investigation is needed to fully understand this behavior. Overall, these results suggest that scaling the teacher model generally enhances the performance of encoder-only and encoder-decoder models.

\textbf{Impact of scaling student model} Row 7-8 of Table \ref{tab: table 4} illustrates the impact of scaling the student model on the performance of learned sparse retrieval (LSR). Both encoder-decoder and decoder-only backbones showed improvements as the student model parameters were scaled. Specifically, the nDCG@10 of the encoder-decoder backbone increased to 70.7 on DL2020 and 73.3 on DL2019, while the MRR@10 rose to 37.8 on MS MARCO. Similarly, the nDCG@10 of the decoder-only backbone increased to 60.9 on DL2020 and 73.1 on DL2019, with the MRR@10 rising to 36.9 on MS MARCO. Notably, the improvement observed for decoder-only backbones was significantly larger when scaling the student model compared to scaling the teacher model.

\section{Related Work}
\subsection{Dense Retrieval}
Dense retrieval encode queries and documents into latent dense representations. Early work on dense retrieval models used encoder transformers, but more recent efforts have been exploring other transformer architectures. Sentence T5~\cite{ni2021sentencet5} compares different transformer architectures (encoder-decoder, encoder-only, decoder-only) for dense sentence embeddings and finds that the encoder-decoder architecture achieves the best performance on semantic textual similarity (STS) benchmarks.  Ni et al.~\cite{ni-etal-2022-large} later extend Sentence T5 (encoder-only) for retrieval tasks and study its effectiveness on different model scales (base, large, XL, XXL). The study suggests that scaling up the T5-encoder could improve the out-of-domain generalizability of dense retrieval. Similarly,  Ma et al.~\cite{ma2023finetuning} developed RepLlama based on Llama---a billion-scale decoder-only language model---and demonstrated the strong in-domain and out-of-domain performance of this model.

\subsection{Learned Sparse Representations}
Learned sparse retrieval offers an efficient alternative to dense retrieval by effectively utilizing traditional inverted indexes, which combines the capabilities of neural methods with the efficiency of classical indexing method. 
DeepCT \cite{dai2019contextaware} and uniCOIL \cite{lin2021brief} use BERT-based model to learn the importance of terms in query and passage. Although they do improve the weighting of terms compared to unsupervised approaches such as BM25, they only weight those terms that are already present in the document, hence lacking term expansions. Therefore, their effectiveness could be limited due to the term mismatch issue. To overcome this term mismatch problem, Mallia et al~\cite{mallia2021learning} improve effectiveness by enriching documents with predicted queries from DocT5Query and proposes a term weighting model that calculates the pairwise loss between relevant and nonrelevant texts with respect to a query. In another approach, Splade~\cite{formal2021splade} directly utilizes language model logits to weight and expand terms in an end-to-end fashion. MacAvaney et al.~\cite{macavaney2020expansion} previously proposed EPIC, which has a similar architecture to Splade but with an MLP query encoder that does not perform query expansion. In a later study, 
Nguyen et al.~\cite{nguyen2023unified} found that under state-of-the-art training configurations, EPIC, without query expansion capability, can be as effective as Splade while being more efficient. This finding is also confirmed by a recent Splade-v3~\cite{lassance2024spladev3} technical report. Unlike dense retrieval, Most of the previous LSR methods has predominantly focused on using small-scale encoder-only transformer models (e.g., DistilBERT, BERT). A concurrent work \cite{doshi2024mistralspladellmsbetterlearned} uses 7 billion model Mistral as the backbone to develop a learned sparse retriever trained on extensive data. To ensure fair comparisons and avoid confounding factors, our study uses the same training data with Splade-DistillMax \cite{lassance2024spladev3}. We address the aforementioned gap in the literature by investigating the performance of LSR across a range of transformer-based architectures, including encoder-only, decoder-only, and encoder-decoder models, as well as larger model scales.

\section{Conclusion}
Our work explored the utility of different transformer-based backbones for LSR. We highlight the difficulty of creating effective sparse representations using LLMs in zero-shot settings. Doing so leads to either the wrong type of term expansion or to a reduction in performance due to an inability to expand. The contribution of our work is to propose a solution for creating sparse representation from the decoder output and investigate the LSR effectiveness across different backbones for LSR.  Our results indicate that incorporating the multi-tokens decoding approach helps to create a more effective sparse representation for the encoder-decoder and decoder-only backbone. 

\section*{Acknowledgments} We thank all reviewers for their feedback. This research was supported by the China Scholarship Council under grant number 202208410053 and project VI.Vidi.223.166 of the NWO Talent Programme (partly) financed by the Dutch Research Council (NWO).
The views expressed in this paper are those of the authors and do not necessarily reflect the views of their institutions or sponsors. 

\newpage
\bibliographystyle{splncs04}
\bibliography{reference}

\appendix 
\clearpage\section*{Appendix}

\begin{table}[h!]
    \centering
    \caption{Teacher performance comparison on the MS MARCO training set.}
    \begin{tabular}{ccc} 
    \toprule
     & nDCG@10 & RR@10 \\ 
     \cmidrule(r){2-3}
    MiniLM-L-6-v2 &  44.3&  37.5\\ 
     Rankllama-13B&  47.5&  40.3\\
    \bottomrule
    \end{tabular}
    \label{tab: table 5}
\end{table}

\begin{figure*}[h!]
    \centering
    \begin{subfigure}{0.49\textwidth}
        \includegraphics[width=\linewidth]{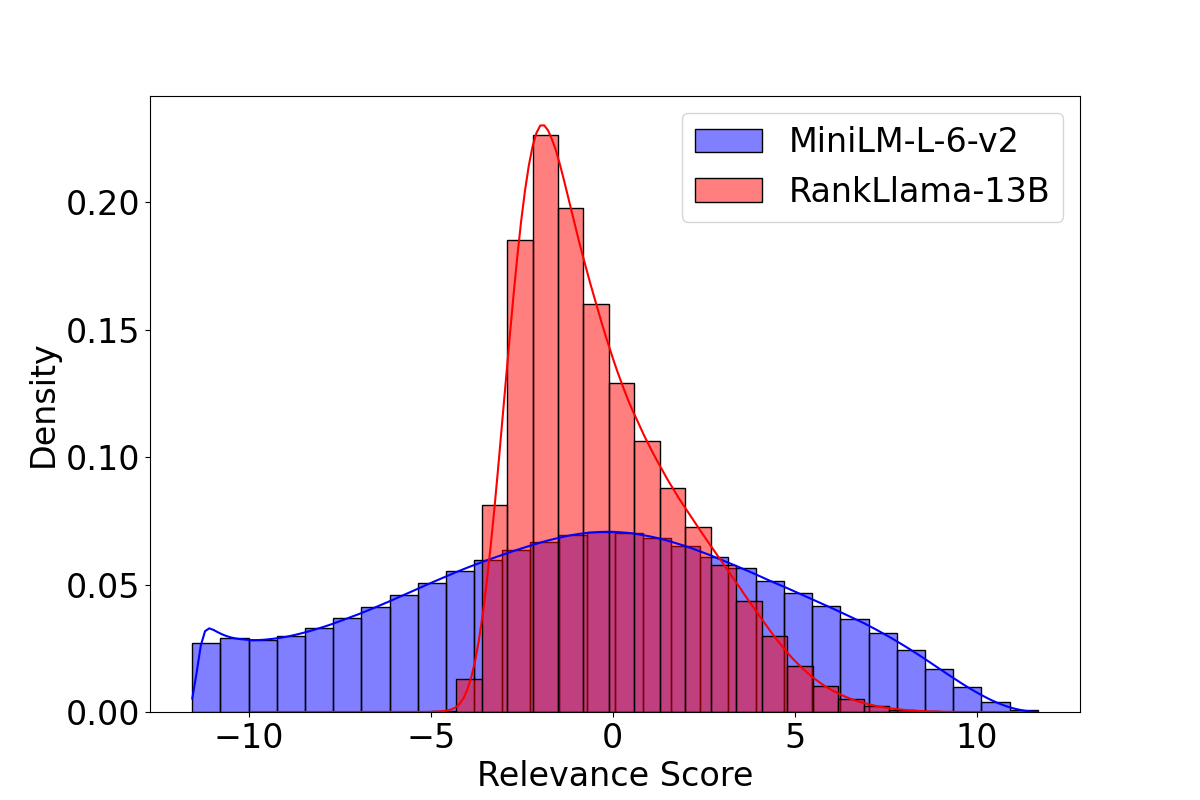}
        \caption{Score distributions of MiniLM-L-6-v2 and RankLLama-13B.}
    \end{subfigure}
    \begin{subfigure}{0.49\textwidth}
        \includegraphics[width=\linewidth]{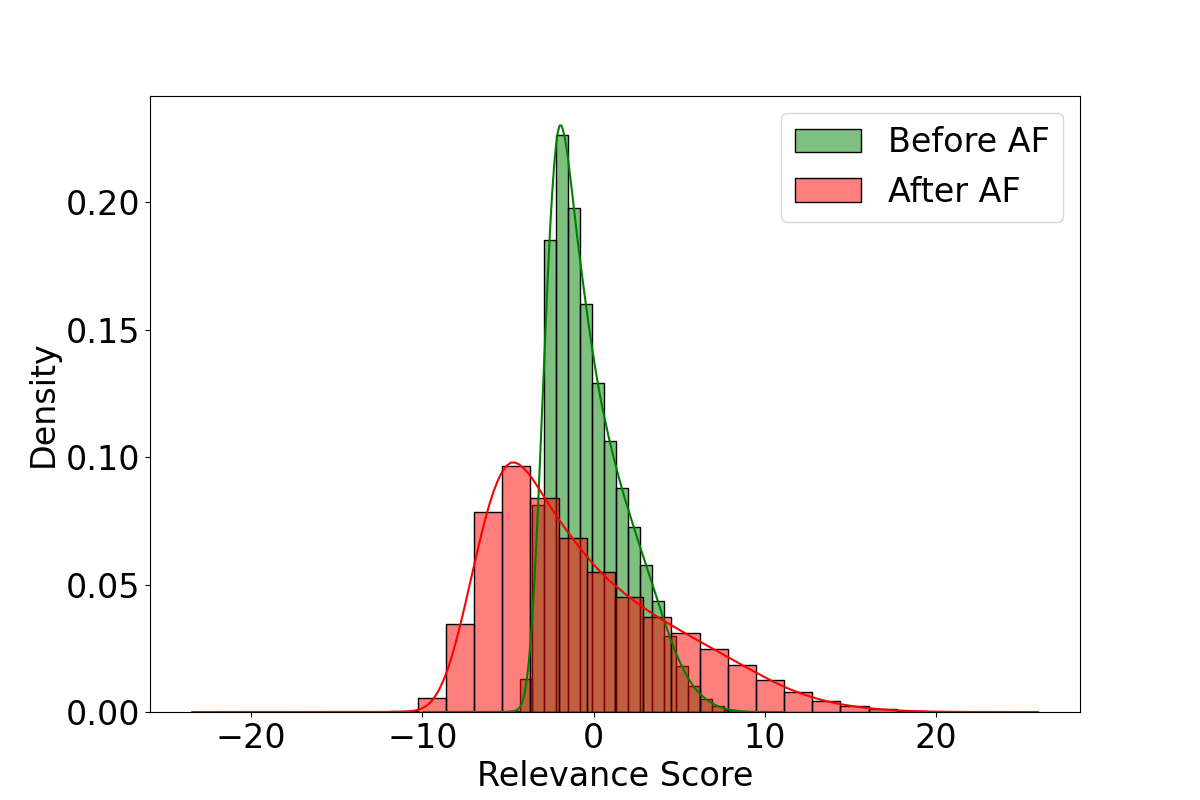}
        \caption{RankLLama-13B scores before and after affine transformation.}
    \end{subfigure} 
    \caption{Score distributions of the two teacher models on the MS MARCO training set.   (a) RankLLama-13B exhibits a sharper distribution than MiniLM-L-6-v2. (b) We apply an affine transformation to align the mean and standard deviation distribution of RankLama-13B with that of MiniLM-L-6-v2.}
    \label{fig: 3}
\end{figure*}

\begin{table*}[htb!]
  \caption{Comparison of LSR effectiveness across different backbones without fine-tuning on MS MARCO dev. \(\times\) results are from \cite{mallia2021learning}; A \(\dagger\) indicates paired t-test \(p < 0.05\). (\(\dagger\) indicates test between term expansion and non-expansion.}
    \label{tab: table 6}
      \centering
    \begin{tabular}{lllcl}
         \toprule
        \textbf{Model}  & \textbf{Weighting} & \textbf{Expansion} & \textbf{RR@10} & \textbf{R@1k} \\
          \cmidrule(r){1-5}
          
    (1) Enc.$_{\text{Distilbert-base-uncased}}$& Yes& No & \phantom{0}0.930 & 40.4\\
    (2) EncDec.MultiTokens$_{\text{FlanT5-base}}$ & Yes & No & 0 &0 \\
    (3) EncDec.MultiTokens$_{\text{FlanT5-xl}}$ & Yes & No & \phantom{0}3.67$^{\dagger}$ & 34.0$^{\dagger}$\\
    (4) Dec.MultiTokens$_{\text{OPT-350M}}$  & Yes&No& 13.2& 50.9 \\
    (5) Dec.MultiTokens$_{\text{OPT-2.7B}}$  & Yes&No& 13.7& 71.4 \\
    (6) Dec.MultiTokens$_{\text{OPT-6.7B}}$  & Yes&No& 12.5&70.5 \\
     \bottomrule
    \end{tabular}
\end{table*}

\end{document}